\documentclass{rspublic}
\usepackage{dsfont}
\usepackage[dvips]{graphicx}
\usepackage{amssymb,amsfonts}
\usepackage{amsmath,exscale,amsthm,amscd}

\begin{document}

\newcommand{\half}{\mbox{$\textstyle \frac{1}{2}$}}
\newtheorem{defi}{Definition}

\title[Dequantisation of the Dirac Monopole]{Dequantisation of the
Dirac Monopole}

\author[Brody]{Dorje~C.~Brody}

\affiliation{Department of Mathematics, Imperial College London,
London SW7 2BZ, UK}

\date{\today}
\maketitle

\begin{abstract}{magnetic monopole; sheaf cohomology; \\
topological quantisation; spin structure; Dirac operator} Using a
sheaf-theoretic extension of conventional principal bundle theory,
the Dirac monopole is formulated as a spherically symmetric model
free of singularities outside the origin such that the charge may
assume arbitrary real values. For integral charges, the construction
effectively coincides with the usual model. Spin structures and
Dirac operators are also generalised by the same technique.
\end{abstract}

\maketitle

\section{Introduction}
\label{sec:1}

In his classical paper on quantisation of magnetic poles Dirac
(1931) remarked that ``Non-euclidean geometry and non-commutative
algebra, which were at one time considered to be purely fictions of
the mind and pastimes for logical thinkers, have now been found to
be very necessary for the description of general facts of the
physical world. It seems likely that this process of increasing
abstraction will continue in the future and that advance in physics
is to be associated with a continual modification and generalisation
of the axioms at the base of the mathematics rather than with a
logical development of any one mathematical scheme on a fixed
foundation.'' In accordance with this principle, the present paper
further extends the work of Dirac by exploring the `dequantisation'
of  magnetic poles.

Diverse and numerous versions of the magnetic pole construction and
the associated charge quantisation condition of Dirac (1931, 1948)
have appeared in the literature, but the basic model can be most
concisely and accurately formulated in terms of the Hopf fibration
(Wu \& Yang 1975; Trautman 1977; Yang 1977). From a mathematical
viewpoint, the rationale is as follows. The monopole potential is
modelled as a connection $A$ on a nontrivial principal $U(1)$-bundle
$P$ over a subset of Minkowski space. A charged matter field
interacting with the monopole is accordingly modelled as a section
of a vector bundle $E_n$ associated with $P$ via the representation
$\rho(\re^{{\rm i}\theta}) \psi = \re^{{\rm i}n\theta} \psi$ of
$U(1)$ on a complex vector space. Since $A$ then induces the
connection $nA$ on $E_n$, the integer $n$ can be identified with the
electric charge of the matter field (see Urbantke 2003 for a
detailed account of the description of a magnetic monopole in the
language of the Hopf fibration).

However, following the spirit of Dirac, if we  describe the monopole
by a more general mathematical scheme, then the interaction of
matter fields and magnetic poles \textit{with arbitrary real
charges} can also be modelled in a consistent manner. The present
paper introduces a new sheaf-theoretic framework permitting an
explicit construction of arbitrarily charged magnetic monopoles.
This framework is likewise applied to generic $U(1)$-bundles and
also yields, as a by-product, the notion of a quasispin structure
defined on arbitrary space-times. The results suggest that
topological quantisation in general can be viewed from a more
flexible perspective.

Although magnetic monopoles have not been observed experimentally,
one important physical consequence of the present model is that
their detection would not necessarily imply the quantisation of
electric charge. Likewise, an observed violation of charge
quantisation would not necessarily imply the nonexistence of
magnetic monopoles. Furthermore, grand unified theories require the
existence of magnetic monopoles, and, according to conventional
field theory described in terms of manifolds this necessarily
implies charge quantisation, hence, the detection of nonintegral
electric charge would indicate that the present scheme, based on
sheaves rather than manifolds, may be physically more realistic.
Also, a sheaf construction permits the global description of fermion
fields on Lorentzian manifolds possessing no conventional spin
structure, as an alternative to the cumbersome machinery of K\"ahler
fermions.

The paper is organised as follows. In \S\ref{sec:2} we present a
very  brief sketch of sheaf theory for the benefit of readers less
acquainted with the subject. This is intended to provide the bare
minimum of  information necessary for following the ensuing
discussion; for further details on sheaf theory, see, e.g., Bredon
(1997), Kultze (1970), or Wells (2008). In \S\ref{sec:3} we develop
the basic mathematical machinery used in later sections. The key
idea here is the construction of a principal ${\mathcal G}$-sheaf
bundle which generalises the conventional notion of a principal
bundle. In \S\ref{sec:4} we prove that equivalence classes of
principal $G$-bundles can, under certain hypotheses, be mapped
injectively into equivalence classes of ${\mathcal G}$-sheaf
bundles. This implies that the Dirac sheaf bundle constructed in
\S\ref{sec:5} does indeed constitute a generalisation of the
conventional Dirac monopole. The spherically symmetrical connection
and curvature of the Dirac sheaf bundle are constructed in
\S\ref{sec:6}, demonstrating that the magnetic charge of this model
can assume arbitrary real values. The interaction of the generalised
monopole with a charged matter field is considered in \S\ref{sec:7}.
Dequantisation of more general $U(1)$ bundles is considered in
\S\ref{sec:8}, with particular attention to gravitational and
electromagnetic instantons. In \S\ref{sec:9} the basic machinery
developed earlier is applied to spin structures and Dirac operators.
In \S\ref{sec:10} the paper concludes with a brief discussion of
possible implications in diverse areas of physics.

\section{Elements of sheaf theory}
\label{sec:2}

The concept of a sheaf over a manifold $X$ provides a way of
interpolating local data and global data on X. We begin with the
definition of a presheaf. A \textit{presheaf} ${\mathfrak F}$ on $X$
is a functor assigning, to each open $U\subset X$, a group $F(U)$,
abelian or otherwise, such that for each $V\subset U$ the
restriction map $r^U_V:\, F(U) \to F(V)$, $r^U_U=1$, defines a
homomorphism and such that $r^V_W r^U_V =r^U_W$ for $W\subset V
\subset U$. An element $\sigma\in F(U)$ is referred to as a
\textit{section} of $F(U)$ over $U$. The restriction of $\sigma\in
F(U)$ on $V\subset U$ is thus given by $\sigma|_{V}=r^U_V(\sigma)$.

The sections $\sigma\in F(U)$ and $\tau\in F(V)$ are said to be
\textit{equivalent} at $x\in U\cap V$ if there is a neighbourhood
$W$ of $x$ such that $r^{U}_{W}(\sigma)= r^{V}_{W}(\tau)$. The
equivalence class containing $\sigma\in F(U)$ is called the
\textit{germ} of $\sigma$ at $x$, and the set of all such germs for
any fixed $x$ is denoted by ${\mathcal F}_x$.

The disjoint union ${\mathcal F}$ of all the sets ${\mathcal F}_x$
provides local information about the structure of ${\mathfrak F}$.
However, information concerning global structure has been lost,
since we have discarded relations between the ${\mathcal F}_x$ for
varying $x$. To retrieve some global structure, we introduce a
topology in the following manner. For a fixed $\sigma\in F(U)$ the
set of all germs $\sigma_x\in{\mathcal F}_x$ for $x\in U$ is taken
to be an open set in ${\mathcal F}$, and the topology of ${\mathcal
F}$ is defined as that generated by these open sets.

The projection $\pi:{\mathcal F}\to X$ mapping ${\mathcal F}_x$ into
$x$ has the property that for any point $t\in{\mathcal F}$ with
$\pi(t) = y$ there is a neighbourhood $N =\{\sigma_x|x\in U\}$ for
$\sigma\in F(U)$ and $\sigma_y = t$ such that the restriction
$\pi|_N$ is a homeomorphism onto a neighbourhood of $y$. These ideas
can be summarised as follows:

\begin{defi} {\rm A \textit{sheaf} of groups on
$X$ is a pair $({\mathcal F},\pi)$ such that
\begin{itemize}
\item[i)] ${\mathcal F}$ is a topological space (in general, not
Hausdorff);
\item[ii)] $\pi$ is a local homeomorphism of ${\mathcal F}$ onto
$X$;
\item[iii)] each ${\mathcal F}_x=\pi^{-1}(x)$, $x\in X$, is a group
called the \textit{stalk} of ${\mathcal F}$ at $x$; and
\item[iv)] the group operations are continuous with respect to the
relative topology on the subset ${\mathcal F}\vartriangle {\mathcal
F} = \{(f,f') \in {\mathcal F}\times{\mathcal F}|\,\pi(f) =
\pi(f')\}$ of ${\mathcal F}\times{\mathcal F}$.
\end{itemize}
}\label{def:1}
\end{defi}

For $U\subset X$, a continuous map $\sigma:U\to{\mathcal F}$ such
that $\pi \sigma(x) = x$ is called a \textit{section} of ${\mathcal
F}$ over $U$. The totality of such $\sigma$ will be denoted
${\boldsymbol\Gamma}(U,{\mathcal F})$. Note that every element of
the group $F(U)$ specified by the presheaf functor naturally
determines an element of ${\boldsymbol\Gamma}(U,{\mathcal F})$, but
the converse is only true locally.

Some of the key properties of a sheaf are as follows. If $\{U_\alpha
\}_{\alpha\in\Lambda}$ is an open covering of an open set $U\subset
X$, and if $\sigma,\sigma'\in F(U)$ are such that $\sigma
|_{U_\alpha} = \sigma'|_{U_\alpha}$ for all $\alpha$, then $\sigma=
\sigma'$. Furthermore, if $\sigma_\alpha\in F(U_\alpha)$ are such
that $\sigma_\alpha|_{U_\alpha\cap U_\beta} = \sigma_\beta
|_{U_\alpha\cap U_\beta}$ for all $\alpha,\beta$, then there exists
an element $\sigma\in F(U)$ with $\sigma|_{U_\alpha}=\sigma_\alpha$
for all $\alpha$. The first property implies that if the
restrictions of a pair of sections always agree, then the two
sections are identical---thus a section over $U$ is determined by
the totality of its restrictions to subsets of $U$. The second
property, somewhat complementary to the first, implies that if pairs
of sections always agree on their overlapping regions then global
section can be constructed from the local data---thus a section over
$U$ may be assembled from consistent local sections on subsets of
$U$.

A sheaf ${\mathcal F}$ contains localised information concerning the
topological space $X$. Global information about $X$ can then be
extracted from ${\mathcal F}$ by consideration of exact sequences,
quotients, and so on. Given two sheaves ${\mathcal F}$  and
${\mathcal F}'$ over $X$, a sheaf homomorphism $\phi:{\mathcal F}
\to {\mathcal F}'$ is a continuous map such that the stalk map
$\phi_x:=\phi|_{{\mathcal F}_x}$ is a homomorphism of ${\mathcal
F}_x$ into ${\mathcal F}'_x$ for each $x\in X$. A sequence of sheaf
homomorphisms of the form
\[
\begin{CD}
0  @>>> {\mathcal F} @>>> {\mathcal G} @>>> {\mathcal H} @>>> 0
\end{CD}
\]
such that the corresponding sequence of stalk maps is exact for all
$x\in X$ is called a short exact sequence of sheaves. Evidently,
exactness is a local property. Given a short exact sequence of
sheaves, the induced sequence
\[
\begin{CD}
0  @>>> {\boldsymbol\Gamma}(X,{\mathcal F}) @>>>
{\boldsymbol\Gamma}(X,{\mathcal G}) @>>>
{\boldsymbol\Gamma}(X,{\mathcal H}) @>>> 0
\end{CD}
\]
is exact at ${\boldsymbol\Gamma}(X,{\mathcal F})$ and
${\boldsymbol\Gamma}(X,{\mathcal G})$, but in general not at
${\boldsymbol\Gamma}(X,{\mathcal H})$. That is to say, the local
exactness of a sequence does not imply exactness with respect to the
global sections over $X$. The measure of inexactness at
${\boldsymbol\Gamma}(X,{\mathcal H})$ can then be characterised by
cohomology.

Recall that in the cohomology theories of $X$ one computes
$H^i(X,G)$ where $G$ is an abelian group. In sheaf cohomology the
coefficients are not elements of a fixed group $G$ but are, rather,
local sections of some sheaf ${\mathcal F}$ over $X$. More
precisely, let ${\mathcal U}=\{U_\alpha\}_{\alpha\in\Lambda}$ be an
open covering of $X$. For any  $U=(U_1,\ldots,U_{q+1})$ such that
$V_U=U_1\cap\cdots\cap U_{q+1}\neq\varnothing$, we define the set of
$q$-cochains by $C^q({\mathcal U},{\mathcal F}) =
\Pi_U{\boldsymbol\Gamma}(V_U,{\mathcal F})$. For any $f\in
C^q({\mathcal U},{\mathcal F})$ define the coboundary operator
$\delta$ by
\begin{eqnarray}
\delta f(V) = \sum_{i=1}^{q+2} (-1)^i r^{V_i}_{V} f(V_i),
\label{eq:2.1}
\end{eqnarray}
where $U_i = (U_1,\ldots,U_{i-1}, U_{i+1}, \ldots,U_{q+1})$ and
$r^{V_i}_{V}$ is the sheaf restriction map. These coboundary
operators define a complex
\[
\begin{CD}
\cdots  @>>> C^{q-1} @>\delta^{q-1}>> C^q @>\delta^q>> C^{q+1} @>>>
\cdots
\end{CD}
\]
and the cohomology groups of this complex are then defined in the
usual manner:
\begin{eqnarray}
H^q({\mathcal U},{\mathcal F}) = \frac{{\rm Ker}\,\delta^q}{{\rm Im}
\, \delta^{q-1}}. \label{eq:2.2}
\end{eqnarray}
Passing to a direct limit over progressively finer coverings, we
obtain the sheaf cohomology groups $H^q(X,{\mathcal F})$.

\section{Basic machinery}
\label{sec:3}

The reader will notice that the definitions in this section run
parallel, \textit{mutatis mutandis}, to the conventional definitions
in the theory of fibre bundles. The ${\mathcal G}$-sheaf plays the
role of a trivial and the ${\mathcal G}$-sheaf bundle that of a
generally nontrivial fibre bundle.

\begin{defi} {\rm Let $X$ and $F$ be topological spaces and
${\mathcal G}=(G,{\tilde\pi},X)$ a sheaf of groups over $X$. A
\textit{${\mathcal G}$-sheaf} over $X$ is a triple ${\mathcal
F}=(F,\pi,X)$ such that
\begin{itemize}
\item[i)] $\pi$ is a local homeomorphism of $F$ onto $X$;
\item[ii)] for each $x \in X$, the stalk ${\mathcal G}_x$ operates
(by left action) upon ${\mathcal F}_x = \pi^{-1}(x)$; and
\item[iii)] if $F\vartriangle G = \{(f,g) \in F \times G|\,\pi(f) =
{\tilde\pi}(g)\}$ is equipped with the relative topology in $F\times
G$, then the mapping $k:\ F \vartriangle G \to F$ defined by
$k(f,g)= gf$ is continuous.
\end{itemize}
} \label{def:2}
\end{defi}

For any subset $A\subset X$, the obvious restriction maps
define a ${\mathcal G}|_{A}$-sheaf ${\mathcal F}|_{A}$ over $A$. In
the sequel, when $A$ is clearly understood, we shall, for brevity,
use the term ${\mathcal G}$-sheaf in place of ${\mathcal
G}|_{A}$-sheaf. We call ${\mathcal F}$ a \textit{principal
${\mathcal G}$-sheaf} if ${\mathcal F} = {\mathcal G}$ and
${\mathcal G}$ operates by left translation.

\begin{defi} {\rm Given two ${\mathcal G}$-sheaves ${\mathcal F} =
(F,\pi,X)$ and ${\mathcal F}' = (F',\pi',X)$ over $X$, a continuous
map $\phi:F \to F'$ will be called a \textit{${\mathcal G}$-sheaf
map} of ${\mathcal F}$ into ${\mathcal F}'$ provided
\begin{itemize}
\item[i)] $\pi'\phi = \pi$; and
\item[ii)] for each $x \in X$, the induced map $\phi_x: {\mathcal F}_x
\to {\mathcal F}'_x$ satisfies
\begin{eqnarray}
\phi_x(g_x f_x) = g_x \phi_x(f_x)  \label{eq:3.1}
\end{eqnarray}
for all $f_x \in {\mathcal F}_x$ and $g_x \in {\mathcal G}_x$.
\end{itemize}
}\label{def:3}
\end{defi}

Again, if $A \subset X$, then a ${\mathcal G}|_A$-sheaf map will,
for brevity, be called a ${\mathcal G}$-sheaf map when the
restriction is clearly understood. A \textit{${\mathcal G}$-sheaf
isomorphism} is a bijective ${\mathcal G}$-sheaf map.

\begin{defi}
{\rm A \textit{${\mathcal G}$-sheaf bundle} ${\mathcal B}$ over a
topological space $X$ is defined by the following data:
\begin{itemize}
\item[i)] an open covering ${\mathcal U} = \{U_\alpha\}_{\alpha
\in\Lambda}$ of $X$;
\item[ii)] for each $\alpha \in\Lambda$, a ${\mathcal G}$-sheaf
${\mathcal F}_\alpha$ over $U_\alpha$; and
\item[iii)] for each nonempty intersection $U_\alpha \cap U_\beta
\neq \varnothing$, a ${\mathcal G}$-sheaf isomorphism
\begin{eqnarray}
T_{\alpha\beta} :\ {\mathcal F}_\beta|_{U_\alpha \cap U_\beta} \to
{\mathcal F}_\alpha|_{U_\alpha \cap U_\beta}  \label{eq:3.2}
\end{eqnarray}
such that the cocycle condition
\begin{eqnarray}
T_{\alpha\beta}T_{\beta\gamma} = T_{\alpha\gamma} \label{eq:3.3}
\end{eqnarray}
is satisfied on $U_\alpha\cap U_\beta\cap U_\gamma$.
\end{itemize}
}\label{def:4}
\end{defi}

We call ${\mathcal B}$ a \textit{principal ${\mathcal G}$-sheaf
bundle} if each ${\mathcal F}_\alpha$ is a principal ${\mathcal
G}$-sheaf. Two ${\mathcal G}$-sheaf bundles ${\mathcal B}$ and
${\mathcal B}'$ over $X$, defined in terms of the respective open
coverings ${\mathcal U}$ and ${\mathcal U}'$, are
\textit{equivalent} provided that the ${\mathcal G}$-sheaf bundles
${\bar{\mathcal B}}$ and ${\bar{\mathcal B}}'$ induced by the
respective restriction maps on some common refinement ${\bar{\mathcal U}}
= \{U_\alpha\}_{\alpha \in\Lambda}$ of ${\mathcal U}$ and ${\mathcal
U}'$ are such that there exist ${\mathcal G}$-sheaf isomorphisms
$T_\alpha: {\mathcal F}_{\alpha} \to {\mathcal F}'_{\alpha}$ for
each $\alpha \in\Lambda$ satisfying
\begin{eqnarray}
{\bar T}_{\alpha\beta} = T_\alpha^{-1} {\bar T}'_{\alpha\beta} T_\beta
\label{eq:3.4}
\end{eqnarray}
for all $\alpha, \beta \in\Lambda$.

Remark: Of academic interest is the fact that, given a ${\mathcal
G}$-sheaf bundle ${\mathcal B}$, one can, from the presheaf defined
by local sections of ${\mathcal B}$, define a ${\mathcal G}$-sheaf
${\mathcal F}_{\mathcal B}$ which in turn uniquely determines the
structure of ${\mathcal B}$ up to equivalence (Kashiwara \& Schapira
2006). However, this fact will not be required in the sequel.

\begin{defi}\label{def:5}
{\rm Let $X$ be a smooth manifold, $G$ a Lie group, and $H$ a closed
central subgroup of $G$. For each open subset $U$ of $X$, let
${\mathit\Gamma}(U,G)$ denote the totality of smooth maps $U \to G$,
which forms a group under pointwise multiplication, i.e. $(\sigma
\sigma')(x) := \sigma(x) \sigma'(x)$. Let ${\mathit\Gamma}_c(U,H)$
denote the totality of constant maps $U \to H$, which forms a
central subgroup of ${\mathit\Gamma}(U,G)$. Assign to each open $U
\subset X$ the quotient group
\begin{eqnarray}
{\mathfrak G}_H^G(U) := {\mathit\Gamma}(U,G)/{\mathit\Gamma}_c(U,
H). \label{eq:3.5}
\end{eqnarray}
With the obvious restriction maps, this defines a presheaf; denote
by ${\mathcal G}_H^G(X)$ the corresponding sheaf of groups over $X$,
regarded as a principal ${\mathcal G}(X)$-sheaf, and its restriction
to a subset $A$ of $X$ by ${\mathcal G}_H^G(A)$.  A principal
${\mathcal G}$-sheaf bundle over $X$ such that ${\mathcal F}_\alpha
= {\mathcal G}_H^G(U_\alpha)$ for some open covering
$\{U_\alpha\}_{\alpha \in \Lambda}$ of $X$ will, for brevity, be
called a \textit{$G_H$-bundle}.}
\end{defi}

\textit{Example 1}. With the notation of Definition~\ref{def:4}, let
$(P,\pi, X)$ be a principal $G$-bundle, and $\{U_\alpha\}_{\alpha
\in \Lambda}$ an open covering of $X$ by trivialising
neighbourhoods. On each nonempty intersection $U_\alpha \cap
U_\beta$ the bundle transition function $g_{\alpha\beta} \in
{\mathit\Gamma}(U_\alpha \cap U_\beta,G)$ defines an element of
${\mathfrak G}_H^G(U_\alpha \cap U_\beta)$ and hence a continuous
section ${\tilde g}_{\alpha\beta}$ of the sheaf ${\mathcal
G}_H^G(U_\alpha \cap U_\beta)$. The germs of ${\tilde
g}_{\alpha\beta}$, acting by right multiplication, provide a
${\mathcal G}$-sheaf map $T_{\alpha\beta}: {\mathcal F}_\beta
|_{U_\alpha \cap U_\beta} \to {\mathcal F}_\alpha|_{U_\alpha \cap
U_\beta}$, which is obviously bijective, i.e. is a ${\mathcal
G}$-sheaf isomorphism. Moreover, the cocycle condition
(\ref{eq:3.3}) follows immediately from the corresponding cocycle
condition for the bundle transition functions. Thus, we obtain a
$G_H$-bundle ${\mathcal F}_P(H)$ over $X$.

\section{Classification of $G_H$-bundles}
\label{sec:4}

The following result shows that under a certain simple hypothesis
the conventional principal $G$-bundles may be identified with a
subset of the $G_H$-bundles over $X$. Throughout the sequel, we
shall assume that $X$ is paracompact, i.e. that every open covering
of $X$ has a locally finite subcovering.

\begin{proposition} \label{prop:1}
For fixed $H$, the correspondence $P \to {\mathcal F}_P(H)$ induces
an injective map of equivalence classes of principal $G$-bundles
into equivalence classes of ${\mathcal G}$-sheaf bundles over $X$,
provided that the \v{C}ech cohomology group $\check{H}^1(X,H)=0$.
\end{proposition}

We first prove three lemmata.

\begin{lemma} \label{lem:2.2}
Let $G$ be an arbitrary group. A mapping $\phi: G \to G$ which
commutes with all left translations is a right translation.
\end{lemma}

Proof. Given that $\phi(g_1 g_2) = g_1 \phi(g_2)$ for all $(g_1,
g_2) \in G\times G$, let $g_2 = e$. Then $\phi(g_1) = g_1 \phi(e)$
for all $g_1 \in G$. \hspace*{\fill} $\square$ \vspace{0.2cm}

Recall that ${\boldsymbol\Gamma}(V, {\mathcal F})$ denotes the
totality of continuous sections of a sheaf ${\mathcal F}$ over a
subset $V\subset X$.

\begin{lemma} \label{lem:2.3}
Let $U$ be an open subset of the topological space $X$, $G$ a group,
${\mathcal F}$ a principal ${\mathcal G}$-sheaf over $X$ and $T:
{\mathcal F}|_U \to {\mathcal F}|_U$ a ${\mathcal G}$-sheaf
isomorphism. Then, for each point $x \in U$, there exists a
neighbourhood $V_x$ of $x$ in $U$ and a $g \in {\boldsymbol\Gamma}
(V_x, {\mathcal F})$ such that $T\sigma(y) = \sigma(y)g(y)$ for
every section $\sigma \in {\boldsymbol\Gamma}(V_x, {\mathcal F})$
and all $y \in V_x$.
\end{lemma}

Proof. By Lemma~\ref{lem:2.2} and condition ii) of
Definition~\ref{def:3}, for each $x \in U$ there exists a $g_0(x)
\in {\mathcal F}_x$ such that $T_x f_x = f_x g_0(x)$ for all $f_x
\in {\mathcal F}_x$. Now, fix $x_0 \in U$, and choose any $f_{x_0}
\in {\mathcal F}_{x_0}$ and any section $\sigma \in
{\boldsymbol\Gamma}(V_{x_0},{\mathcal F})$ with $V_{x_0}$ open in
$U$, $x_0 \in V_{x_0}$ and $\sigma(x_0) = f_{x_0}$. Since $T$ is
continuous and stalk-preserving, $T\sigma(y) = \sigma(y) g_0(y)$ ($y
\in V_{x_0}$) defines a continuous section of ${\mathcal F}$, and by
the continuity of the group operations $g_0(y) = \sigma(y)^{-1}
T\sigma(y)$ $(y \in V_{x_0})$ is also a continuous section. For any
other section $\sigma'(y) \in {\boldsymbol\Gamma}(V_{x_0}, {\mathcal
F})$, writing $\sigma'(y) = f'_y$ we have $T\sigma'(y) = T_y f'_y =
f'_y g_0(y)= \sigma'(y) g_0(y)$. Hence, $g_0(y) \in
{\boldsymbol\Gamma} (V_{x_0},{\mathcal F})$ has the desired
property. \hspace*{\fill} $\square$ \vspace{0.2cm}

\begin{lemma} \label{lem:2.4}
Let ${\mathcal F} = (F,\pi, X)$ and ${\mathcal F}' = (F, \pi', X)$
be sheaves, $\phi: {\mathcal F} \to {\mathcal F}'$ an epimorphic
sheaf map and $\sigma' \in {\boldsymbol\Gamma}(X, {\mathcal F}')$.
Then, for any $x \in X$, there exists a neighbourhood $V_x$ of $x$
and a $\sigma \in {\boldsymbol\Gamma} (V_x, {\mathcal F})$ such that
$\phi \sigma = \sigma'|_{V_x}$.
\end{lemma}

Proof. This is an elementary fact of sheaf theory (Kultze 1970,
Hilfsatz 3.4).  \hspace*{\fill} $\square$ \vspace{0.2cm}

Proof of Proposition~\ref{prop:1}. Clearly, equivalent principal
bundles give rise to equivalent sheaf bundles. Conversely, let
$(P,\pi,X)$ and $(P',\pi',X)$ be principal $G$-bundles defined in
terms of trivialising open coverings ${\mathcal U}$ and ${\mathcal
U}'$, respectively. Choosing, if necessary, a common refinement, we
may suppose that ${\mathcal U} = {\mathcal U}' =
\{U_{\alpha}\}_{\alpha\in\Lambda}$. Then, with the foregoing
notation, ${\mathcal F}_P(H) \sim {\mathcal F}'_P(H)$ iff there
exist ${\mathcal G}$-sheaf isomorphisms $T_{\alpha}: {\mathcal
F}_{\alpha} \to {\mathcal F}'_{\alpha}$, $\alpha \in\Lambda$, such
that
\begin{eqnarray}
T'_{\alpha\beta} = T_\alpha T_{\alpha\beta} T_{\beta}^{-1}
\label{eq:4.1}
\end{eqnarray}
on $U_{\alpha} \cap U_\beta$ ($\alpha, \beta \in\Lambda$). The
${\mathcal G}$-sheaves ${\mathcal F}_{\alpha}$ and ${\mathcal
F}'_{\alpha}$ are identical, both arising from the presheaf functor
${\mathfrak F}_\alpha(V) = {\mathit\Gamma}(V,
G)/{\mathit\Gamma}_c(V, H)$, $V$ open in $U_{\alpha}$. Let
${\mathcal G}_{\alpha}$ denote the sheaf over $U_\alpha$ defined by
the presheaf functor $V\to{\mathit\Gamma}(V,G)$. Then, the presheaf
epimorphism ${\mathit\Gamma}(V,G) \to {\mathfrak F}_{\alpha}(V)$
induces an epimorphic sheaf map $\phi: {\mathcal G}_{\alpha} \to
{\mathcal F}_{\alpha}$, hence, by Lemmata~\ref{lem:2.3} and
\ref{lem:2.4}, for each $x \in U_{\alpha}$ there exists a
neighbourhood $V_x \subset U_{\alpha}$ and a section $g_{\alpha}^{x}
\in {\mathit\Gamma}(V_x, G)$ such that $T_{\alpha}[\sigma_{\alpha}]
= [\sigma_{\alpha}][g_{\alpha}^{x}] =
[\sigma_{\alpha}g_{\alpha}^{x}]$ for all $\sigma_{\alpha} \in
{\mathit\Gamma}(V_x, G)$, where the square brackets denote cosets in
${\mathfrak G}^G_H(V_x)$, regarded as continuous sections in
${\boldsymbol\Gamma}(V_x, {\mathcal F}_\alpha)$. Then, choosing the
covering ${\mathcal U} =\{U_{\alpha}\}_{\alpha\in\Lambda}$
sufficiently fine, we may, by paracompactness, omit the superscript
$x$ and simply write $T_{\alpha}[\sigma_{\alpha}]=
[\sigma_{\alpha}g_{\alpha}]$ for all $\sigma_{\alpha} \in
{\mathit\Gamma}(U_{\alpha},G)$, with square brackets denoting cosets
in ${\mathfrak G}^G_H(U_\alpha)$, again regarded as elements of
${\boldsymbol\Gamma}(U_\alpha, {\mathcal F}_\alpha)$. Hence,
recalling the definition of the $T_{\alpha\beta}$ in Example 1,
relation (\ref{eq:4.1}) implies that
\begin{eqnarray}
[\sigma_{\alpha}g'_{\alpha\beta}] = [\sigma_{\alpha}g_{\beta}^{-1}
g_{\alpha\beta} g_\alpha]   \label{eq:4.2}
\end{eqnarray}
on $U_{\alpha} \cap U_{\beta}$. Since $\sigma_\alpha$ is arbitrary,
this clearly means that
\begin{eqnarray}
g'_{\alpha\beta}(x) = g_\beta^{-1}(x) g_{\alpha\beta}(x) g_\alpha(x)
h_{\alpha\beta} \label{eq:4.3}
\end{eqnarray}
for certain constant functions
\begin{eqnarray}
h_{\alpha\beta}: U_{\alpha} \cap U_\beta \to H. \label{eq:4.4}
\end{eqnarray}
Applying the cocycle condition for principal bundles to both members
of (\ref{eq:4.3}), we deduce that the constants $h_{\alpha\beta}$
also satisfy the cocycle condition, and therefore define an element
of $\check{H}^1(X,H)$. Hence, if the covering $\{U_{\alpha}
\}_{\alpha\in\Lambda}$ is chosen sufficiently fine, then, by
hypothesis, there exists a 0-cochain
$\{h_{\alpha}\}_{\alpha\in\Lambda}$ such that $h_{\alpha\beta} =
h_{\beta}^{-1} h_\alpha$. Defining ${\bar g}_{\alpha}(x) =
h_{\alpha} g_{\alpha}(x)$ we find that (\ref{eq:4.3}) becomes
\begin{eqnarray}
g'_{\alpha\beta}(x) = {\bar g}_\beta^{-1}(x)g_{\alpha\beta}(x) {\bar
g}_{\alpha}(x), \label{eq:4.5}
\end{eqnarray}
which is the condition for equivalence of $P$ and $P'$.
\hspace*{\fill} $\square$ \vspace{0.2cm}

As used in the foregoing proof, Lemmata~\ref{lem:2.3} and
\ref{lem:2.4} show that for a sufficiently fine covering ${\mathcal
U}$ the transition sheaf isomorphisms of a $G_H$-bundle are of the
form
\begin{eqnarray}
T_{\alpha\beta} {\bar\sigma} = {\bar\sigma} {\bar g}_{\alpha\beta}
\label{eq:4.6}
\end{eqnarray}
for fixed ${\bar g}_{\alpha\beta}$ and arbitrary ${\bar\sigma} \in
{\boldsymbol\Gamma}(U_{\alpha} \cap U_\beta, {\mathcal F}_{\beta})$.
Likewise, the ${\mathcal G}$-sheaf isomorphisms $T_\alpha$ in the
definition of ${\mathcal G}$-sheaf bundle equivalence are of the
form
\begin{eqnarray}
T_\alpha {\bar\sigma} = {\bar\sigma}{\bar g}_\alpha \label{eq:4.7}
\end{eqnarray}
for fixed ${\bar g}_{\alpha}$ and arbitrary ${\bar\sigma} \in
{\boldsymbol\Gamma}(U_\alpha,{\mathcal F}_{\alpha})$.

A standard theorem states that equivalence classes of principal
$G$-bundles over $X$ correspond biuniquely to elements of the sheaf
cohomology set $H^1(X,{\mathcal G})$. In view of (\ref{eq:4.6}), the
argument used in proving this theorem (see, e.g., Lawson \&
Michelsohn 1989, Appendix A) can also be applied in a
straightforward manner to yield the following result.

\begin{corollary}
The equivalence classes of $G_H$-bundles over $X$ correspond
biuniquely to elements of the sheaf cohomology set $H^1(X,{\mathcal
F})$. For $H = \{e\}$, this correspondence reduces to the standard
theorem cited above. \label{cor:4.05}
\end{corollary}

If $G$ is abelian, then Proposition~\ref{prop:1} follows more
directly from standard results as follows. Let ${\mathcal H}_c$
denote the sheaf arising from the constant presheaf defined by
$H_c(U) = {\mathit\Gamma}_c(U,H) \sim H$. The short exact sequence
\[
\begin{CD}
0  @>>> {\mathit\Gamma}_c(U,H) @>i>> {\mathit\Gamma}(U,G) @>j>>
{\mathit\Gamma}(U,G)/{\mathit\Gamma}_c(U,H) @>>> 0
\end{CD}
\]
induces a sequence of sheaf maps
\[
\begin{CD}
0  @>>> {\mathcal H}_c @>{\boldsymbol i}>> {\mathcal G}
@>{\boldsymbol j}>> {\mathcal F} @>>> 0
\end{CD}
\]
which is easily seen to be exact. This in turn induces a long exact
cohomology sequence which includes, in particular, the segment
\[
\begin{CD}
\cdots @>>> H^1(X,{\mathcal H}_c) @>{\boldsymbol i}^*>>
H^1(X,{\mathcal G}) @>{\boldsymbol j}^*>> H^1(X,{\mathcal F}) \\
@>{{\boldsymbol \delta}^*}>> H^2(X,{\mathcal H}_c) @>>> \cdots
\end{CD}
\]
For a paracompact $G$-manifold $X$ the sheaf cohomology group
$H^p(X,{\mathcal H}_c)$ is known to be isomorphic to the \v{C}ech
cohomology group $\check{H}^p(X,H)$ as well as the singular
cohomology group $H^p(X,H)$ with coefficients in $H$ (Spanier 1994,
Chapter~6). Hence, if $\check{H}^1(X,H) = 0$, then the exactness of
the above sequence implies that ${\boldsymbol j}^*$ is injective.
Denoting the totality of equivalence classes of principal
$G$-bundles (resp. $G_H$-bundles) by $P_G(X)$ (resp. ${\mathcal
P}_{G_H}(X)$), we note that the diagram
\[
\begin{CD}
P_G(X) @>>> {\mathcal P}_{G_H}(X) \\
@VVV @VVV \\
H^1(X,{\mathcal G}) @>{{\boldsymbol j}^*}>> H^1(X,{\mathcal F})
\end{CD}
\]
where the upper arrow represents the correspondence $P \to {\mathcal
F}_P(X)$ and the vertical arrows the biunique correspondence of
Corollary~\ref{cor:4.05}, is commutative. The result follows.

Relations (\ref{eq:4.6}) and (\ref{eq:4.7}) also imply:

\begin{corollary}
For the special case $H =\{e\}$ {\rm (}the identity of $G${\rm )},
the $G_H$-bundles can be identified with conventional principal
$G$-bundles, and ${\mathcal G}$-sheaf equivalence is merely
conventional bundle equivalence. \label{cor:4.5}
\end{corollary}

Thus, for $H =\{e\}$ the present theory yields nothing new. However,
for $H \neq\{e\}$ and $\check{H}^1(X, H)=0$ we obtain a nontrivial
extension of the conventional theory of principal bundles, as we
shall see, in particular, from the example in the next section,
wherein $H=G$.

Note that the set of data ${\mathcal T}=\{U_\alpha, {\mathcal
F}_\alpha, {\bar g}_{\alpha\beta}\}$ (see Definition~\ref{def:5} and
(\ref{eq:4.6})), subsequently referred to as a \textit{presentation}
of the $G_H$-bundle under consideration, plays a role analogous to
that of a system of local trivialisations in the conventional theory
of principal bundles.

If ${\mathcal U}'$ is a refinement of ${\mathcal U} =
\{U_\alpha\}_{\alpha\in \Lambda}$, then ${\mathcal T}$ induces, in
the obvious manner, a presentation ${\mathcal T}'$ associated with
${\mathcal U}'$. Recalling the definition of ${\mathcal G}$-sheaf
equivalence (see (\ref{eq:3.4})), relations (\ref{eq:4.6}) and
(\ref{eq:4.7}) imply that two presentations ${\mathcal T}$ and
${\mathcal T}'$ associated with the same covering
$\{U_\alpha\}_{\alpha\in \Lambda}$ define equivalent $G_H$-bundles
iff there exist elements ${\bar g}_\alpha \in {\boldsymbol
\Gamma}(U_\alpha, {\mathcal F}_\alpha)$ such that the respective
transition elements ${\bar g}_{\alpha\beta}$ and ${\bar
g}'_{\alpha\beta}$ satisfy
\begin{eqnarray}
{\bar g}'_{\alpha\beta}(x) = {\bar g}_\beta^{-1}(x) {\bar
g}_{\alpha\beta}(x) {\bar g}_\alpha(x) h_{\alpha\beta}
\label{eq:4.8}
\end{eqnarray}
for all $\alpha, \beta \in \Lambda$ and $x \in U_\alpha \cap
U_\beta$.  If the associated open coverings ${\mathcal U}$ and
${\mathcal U}'$ are different, then the condition for equivalence is
given by (\ref{eq:4.8}) with respect to some common refinement
${\mathcal U}''$ of ${\mathcal U}$ and ${\mathcal U}'$.

Another immediate consequence of (\ref{eq:4.6}) is the fact that
$G_H$-bundles, like conventional $G$-bundles, are functorial, i.e. a
$G_H$-bundle ${\mathcal P}$ over $X$ and a continuous map $f:\, Y\to
X$ naturally induce a $G_H$-bundle $f^*({\mathcal P})$ over $Y$,
since the transition sections ${\bar g}_{\alpha\beta}$ on $X$ pull
back to sections on $Y$ which obviously satisfy the cocycle
condition. Moreover, one can prove that if $X$ and $Y$ are compact
Hausdorff spaces and the maps $f$ and $f':\ Y\to X$ are homotopic,
then $f^*({\mathcal P})$ and $f^{\prime*}({\mathcal P})$ are
equivalent (cf. Lawson \& Michelsohn 1990, Appendix A).

\section{Dirac sheaf bundles}
\label{sec:5}

Let $X=S^2$ and $G=H=U(1)$. We represent $S^2$ by the unit sphere in
${\mathds R}^3$, with spherical polar coordinates
$(\theta,\phi)$, ($0\leq\theta\leq\pi$, $0\leq\phi<2\pi$), and
$U(1)$ by $S^1 = \{\re^{{\rm i}\gamma}|\, 0 \leq \gamma < 2\pi\}$.
For the open covering
\begin{eqnarray}
U_1 = \{(\theta,\phi)|\, 0 \leq \theta < {\textstyle\frac{3}{2}}
\pi\}, \quad U_2 = \{(\theta,\phi)|\, \half\pi < \theta \leq\pi\}
\label{eq:5.1}
\end{eqnarray}
of $S^2$, we have $U_1 \cap U_2 = \left\{(\theta,\phi)\left|\,
\frac{1}{2}\pi < \theta < \frac{3}{2}\pi\right. \right\}$. Let
${\mathcal G}$ be the sheaf of smooth $U(1)$-valued functions over
$S^2$. Let ${\mathfrak F}$ denote the presheaf over $S^2$ defined
for each open $V$ by
\begin{eqnarray}
{\mathfrak F}(V) = {\mathit\Gamma}(V, U(1))/{\mathit\Gamma}_c(V,
U(1)), \label{eq:5.2}
\end{eqnarray}
${\mathcal F}$ the associated ${\mathcal G}$-sheaf, and ${\mathcal
F}_\alpha = {\mathcal F}|_{U_\alpha}$ $(\alpha=1,2)$. If $\nu$ is an
arbitrary real number, define a ${\mathcal G}$-sheaf isomorphism
\begin{eqnarray}
T_{12}: {\mathcal F}_2|_{U_1 \cap U_2} \to {\mathcal F}_1|_{U_1 \cap
U_2} \label{eq:5.3}
\end{eqnarray}
by
\begin{eqnarray}
T_{12}(x) f_x = f_x {\bar g}^\nu_{12}(x) \label{eq:5.4}
\end{eqnarray}
for $x = (\theta_0,\phi_0) \subset U_1 \cap U_2$, where ${\bar
g}^{\nu}_{12}$ is the section of ${\mathcal F}|_{U_1 \cap U_2}$
defined locally by
\begin{eqnarray}
g^\nu_{12}(\theta,\phi) = \left[ \re^{-{\rm i}\nu \phi} \right].
\label{eq:5.5}
\end{eqnarray}
Here the square brackets indicate the equivalence class modulo
constant sections $c_h(\theta, \phi) = \re^{{\rm i}h}$. Denote the
$G_H$-bundle so defined by ${\mathcal D}_\nu$. If $\nu = n$ is an
integer, then ${\mathcal D}_n$ is just the principal ${\mathcal
G}$-sheaf bundle ${\mathcal F}_{P_n}(U(1))$ arising from the
conventional $U(1)$-bundle $P_n$ of charge $\frac{1}{2}n$ (in the
appropriate units), in accordance with the construction of
Example~1. In general, for an arbitrary abelian group $H$, the
\v{C}ech cohomology groups are given by
\begin{eqnarray}
\check{H}^q(S^n,H) = \left\{ \begin{array}{cl} H & {\rm if}\ q=0\
{\rm or}\ n, \\ 0 & {\rm otherwise} \end{array} \right.
\label{eq:5.6}
\end{eqnarray}
(see, e.g., Spanier 1994). In particular, we have
$\check{H}^1(S^2,U(1))=0$. Hence, by Proposition~\ref{prop:1}, the
sheaf bundles ${\mathcal D}_n$ are mutually inequivalent. That $\nu
\neq \nu'$ implies ${\mathcal D}_\nu \nsim {\mathcal D}_{\nu'}$ for
arbitrary real $\nu$ and $\nu'$ will be proved in \S\ref{sec:6}
below.

From the foregoing construction, the truth of the following
proposition should be evident.

\begin{proposition} \label{prop:5.1}
Let $g_{\alpha\beta}$ be the transition functions for some
trivialising atlas of a principal $U(1)$-bundle $P$ over the base
space $X$. Then, for any real number $\nu$, the powers
$(g_{\alpha\beta})^\nu$ define the transition isomorphisms of a
$U(1)_{U(1)}$-bundle ${\mathcal P}_\nu$ over $X$. If the \v{C}ech
cohomology group $\check{H}^1(X,U(1)) = 0$, then the principal
$U(1)$-bundles $P_n$ defined by the transition functions
$(g_{\alpha\beta})^n$, if mutually inequivalent, correspond
bi-uniquely with mutually inequivalent $U(1)_{U(1)}$-bundles
${\mathcal P}_n$.
\end{proposition}

The final assertion is merely an application of
Proposition~\ref{prop:1}.

\section{Connections on $G_H$-bundles}
\label{sec:6}

The Lie algebra of $G$ will be denoted by ${\mathfrak g}$ and the
algebra of smooth ${\mathfrak g}$-valued differential forms on a
manifold $M$ by $\Lambda(M,{\mathfrak g})$.

\begin{defi}
{\rm A \textit{connection} ${\mathcal A}$ on a $G_H$-bundle
${\mathcal F}$ given in terms of a presentation $(U_\alpha,
{\mathcal F}_\alpha, {\bar g}_{\alpha\beta})$ is specified by a
family of ${\mathfrak g}$-valued 1-forms $A_\alpha \in
\Lambda^1(U_\alpha,{\mathfrak g})$  satisfying\footnote{For
simplicity of notation we assume that $G$ is a matrix group.}
\begin{eqnarray}
A_\alpha (x) = {\bar g}_{\alpha\beta}(x) A_\beta(x) {\bar
g}_{\beta\alpha} (x) + {\bar g}_{\alpha\beta}(x) \rd {\bar
g}_{\beta\alpha}(x) \label{eq:6.1}
\end{eqnarray}
for all $\alpha, \beta\in \Lambda$ and $x \in U_\alpha \cap U_\beta$.
} \label{def:6}
\end{defi}

Both terms on the right side of (\ref{eq:6.1}) are unambiguously
defined as follows. The value of ${\bar g}_{\alpha\beta}$ at $x \in
U_\alpha \cap U_\beta$ is an element of ${\mathcal F}_{\beta x}$,
represented by $f_{\alpha\beta} \in {\mathfrak F}(V_x) = {\mathfrak
G}^G_H(V_x)$ for some neighbourhood $V_x$ of $x$, and
$f_{\alpha\beta}$ is in turn represented by a section
$g_{\alpha\beta} \in {\mathit\Gamma}(V_x,G)$. The first term on the
right side of (\ref{eq:6.1}) is then defined by $g_{\alpha\beta}(x)
A_\beta(x) g_{\beta\alpha}(x)$, and the second by
$g_{\alpha\beta}(x)\rd g_{\beta\alpha}(x)$. Any two possible choices
$g_{\alpha\beta}$ differ only by a constant factor in $H$, which
affects neither term. Two such connections ${\mathcal A}$ and
${\mathcal A}'$ defined in terms of the respective presentations
${\mathcal T}$ and ${\mathcal T}'$ are \textit{equivalent}, provided
that, for some common refinement $\{U_\alpha\}_{\alpha\in \Lambda}$
of the two respectively associated open coverings ${\mathcal U}$ and
${\mathcal U}'$ with (cf. (\ref{eq:4.8}))
\begin{eqnarray}
{\bar g}'_{\alpha\beta}(x) = {\bar g}_\beta^{-1}(x) {\bar
g}_{\alpha\beta}(x)  {\bar g}_\alpha(x) h_{\alpha\beta},
\label{eq:6.2}
\end{eqnarray}
the relation
\begin{eqnarray}
A'_\alpha (x) = {\bar g}_\alpha^{-1}(x) A_\alpha(x) {\bar g}_\alpha
(x) + {\bar g}_\alpha^{-1}(x) \rd {\bar g}_\alpha (x) \label{eq:6.3}
\end{eqnarray}
holds for all $\alpha\in \Lambda$ and $x \in U_\alpha$. Here the right
side of (\ref{eq:6.3}) is interpreted in the same manner as that of
(\ref{eq:6.1}). By choosing a sufficiently fine covering, the
above-mentioned neighbourhoods $V_x$ may, by paracompactness, be
identified with the trivialising neighbourhoods $U_\alpha$, so that
the representative sections ${g}_{\alpha\beta}$ and ${g}_\alpha$ may
be regarded as elements of ${\mathit\Gamma}(U_\alpha \cap U_\beta,
G)$ and ${\mathit\Gamma}(U_\alpha,G)$, respectively. The
${g}_{\alpha\beta}$ then satisfy the cocycle condition modulo
locally constant sections in $H$, which suffices to ensure the
mutual consistency of the relations (6.1) as the indices $\alpha$
and $\beta$ vary over $\Lambda$.

The \textit{curvature} of such a
connection is defined in the conventional manner, i.e. for each
$U_\alpha$ we have
\begin{eqnarray}
F_\alpha = \rd A_\alpha + A_\alpha \wedge A_\alpha = \rd A_\alpha +
\half [A_\alpha, A_\alpha].    \label{eq:6.4}
\end{eqnarray}
By the same calculation as that used for conventional principal
bundles, one finds, using (6.1), that
\begin{eqnarray}
F_\alpha (x) = {\bar g}_{\alpha\beta}(x) F_\beta (x) {\bar
g}_{\beta\alpha}(x) \label{eq:6.5}
\end{eqnarray}
for $x \in U_\alpha \cap U_\beta$. Furthermore, if $A'_\alpha$ and
$A_\alpha$ are related by (\ref{eq:6.3}), then
\begin{eqnarray}
F'_\alpha(x) = {\bar g}_\alpha^{-1}(x) F_\alpha(x) {\bar g}_\alpha
(x).  \label{eq:6.6}
\end{eqnarray}
If $(P,\pi,X)$ is a principal $G$-bundle and $H$ a central closed
subgroup of $G$, then an ordinary connection $A$ on $P$
clearly gives rise to a connection ${\mathcal A}$ on ${\mathcal
F}_P(H)$ in the sense of Definition~\ref{def:6}, and the
conventional local curvature forms then coincide with those of
${\mathcal A}$ as defined in (\ref{eq:6.4}). By virtue of the
relations (\ref{eq:6.5}) and (\ref{eq:6.6}), characteristic classes
of $G_H$-bundles can be defined and shown to be independent of the
choice of the presentation and the choice of the connection
${\mathcal A}$, as for conventional principal bundles.

We now define a connection on the Dirac sheaf bundle ${\mathcal
D}_\nu$ ($\nu \in {\mathds R}$). Using the notation of
\S\ref{sec:5}, let ${\mathcal T}^\nu$ denote the presentation
$\{U_1,U_2, {\mathcal F}_1, {\mathcal F}_2, {\bar g}^\nu_{12}\}$,
and let
\begin{eqnarray}
A_\alpha =  \half\ri\nu\left[(-1)^{\alpha+1} - \cos \theta\right]
\rd\phi \in \Lambda^1(U_\alpha, {\mathfrak u}(1)) \label{eq:6.7}
\end{eqnarray}
be the corresponding local 1-forms. On $U_1 \cap U_2$ we have
\begin{eqnarray}
A_1 = A_2 + \ri \nu = A_2 + g_{12}^\nu \rd g_{21}^\nu,
\label{eq:6.8}
\end{eqnarray}
in accordance with (\ref{eq:6.1}). The curvature
is given by
\begin{eqnarray}
F = \rd A_\alpha = \half \ri \nu \sin\theta\, \rd\theta \wedge
\rd\phi, \label{eq:6.9}
\end{eqnarray}
and hence the first Chen\footnote{I use the current officially valid
romanisation in place of the still prevalent ``Chern", which was
officially discarded about sixty years ago.} class is $-\nu \rd S$,
where $\rd S$ is the area form on the unit 2-sphere. This shows that
the sheaf bundles ${\mathcal D}_\nu$ for distinct $\nu$ are inequivalent, as
claimed above.

For simplicity of exposition and to emphasise the analogy with the
conventional Hopf fibration $S^1 \to S^3 \to S^2$, the Dirac sheaf
bundle and the corresponding monopole connection have been
constructed over $X = S^2$ as base space. However, with a view to
physical applications, the same construction applies verbatim if the
base space is the subset of Minkowski space defined by ${\tilde
{\mathds R}}^4 := \{ (x^0,x^1,x^2,x^3) |(x^1)^2+ (x^2)^2+(x^3)^2 >
0\}$. Again, expressing $(x^1,x^2,x^3)$ by spherical polar
coordinates, formulae (5.1), (5.2) and (5.6) remain unchanged. Thus,
for arbitrary real $\nu$, one obtains a $U(1)_{U(1)}$-bundle over
${\tilde {\mathds R}}^4$ with field strength (\ref{eq:6.9}).

Remark: According to the conventional bundle picture, there are no
singularity-free gauge potentials $\{A_\alpha\}_{\alpha=1,2}$ with
the properties that (i) their curls are equal to the field, and that
(ii) they are related by gauge transformations on the overlapping
region $U_1 \cap U_2$, unless electromagnetic charges are quantised
(Wu \& Yang 1975, Theorem 3). The above result demonstrates that
this conclusion is not valid in the present generalised model.

\section{Particle fields}
\label{sec:7}

Let $G$ be a Lie group, $H$ a closed subgroup of $G$, and $\rho: G
\to Aut({\mathbb V})$ a representation of $G$ on a (real or complex)
vector space ${\mathbb V}$. Let $X$ be a smooth paracompact
manifold, and for each open $U \subset X$, let ${\mathit\Gamma}(U,
{\mathbb V})$ denote the totality of smooth maps $\tau: U\to
{\mathbb V}$. If $\tau \in {\mathit\Gamma}(U,{\mathbb V})$ and $h\in
H$, define $h\tau$ by $(h\tau)(x) = h(\tau(x))$. Denote by
${\mathit\Gamma}_H(U,{\mathbb V})$ the quotient space of
${\mathit\Gamma}(U,{\mathbb V})$ under this action of $H$. If $U'
\subset U$, then the restriction of ${\mathit\Gamma}(U,{\mathbb V})$
to ${\mathit\Gamma}(U',{\mathbb V})$ obviously commutes with the
action of $H$, and thus one obtains a restriction map $r^U_{U'}$. If
$U''\subset U' \subset U$, then $r^{U'}_{U''} r^U_{U'} = r^U_{U''}$,
hence the system $\{{\mathit\Gamma}_H(U,{\mathbb V}), r^U_{U'}\}$
defines a presheaf ${\mathfrak V}$, and a corresponding sheaf
${\mathcal V}.$\footnote{In this context, all the definitions in
\S\ref{sec:3} are to be interpreted with groups replaced by vector
spaces and homomorphisms by linear transformations.}

Next, for $x \in X$, define a left action of the stalk $({\mathcal
G}^G_H)_x$ of the sheaf ${\mathcal G}^G_H$ (see Definition 5) upon
${\mathcal V}_x$ as follows. For some neighbourhood $U$ of $x$, the
germ ${\bar g}_x \in ({\mathcal G}^G_H)_x$ is represented by an
element ${\bar g}\in {\mathfrak G}^G_H(U)$ and ${\bar g}$ in turn by
a section $g \in {\mathit\Gamma}(U,G)$, while ${\bar{v}}_x$ is
represented by an element ${\bar{v}}_x$ of
${\mathit\Gamma}_H(U,{\mathbb V})$ and ${\bar{v}}_x$ in turn by a
section $v \in {\mathit\Gamma}(U,{\mathbb V})$. Define ${\bar g}_x
{\bar v}_x$ as the element of ${\mathcal V}_x$ represented by the
section $(gv)(x) = \rho(g(x)) v(x)$. One can readily check that this
is independent of the choices of $g$ and $v$.

\begin{defi}
{\rm Let $({\mathcal P},\pi,X)$ be a $G_H$-bundle defined by a
presentation ${\mathcal T} = \{U_\alpha, {\mathcal F}_\alpha, {\bar
g}_{\alpha\beta}\}_{\alpha \in \Lambda}$. A \textit{particle field}
is a system of sections $\{\bar{v}_\alpha \in {\boldsymbol\Gamma}
(U_\alpha,{\mathcal V})\}_{\alpha \in \Lambda}$ satisfying the
condition
\begin{eqnarray}
\bar{v}_\alpha(x) = {\bar g}_{\alpha\beta}(x) \bar{v}_\beta(x)
\label{eq:7.1}
\end{eqnarray}
for every $\alpha, \beta \in \Lambda$ and $x \in U_\alpha \cap
U_\beta$. The totality of such particle fields will be denoted by
${\mathcal V}_\rho({\mathcal P})$.} \label{def:7}
\end{defi}

Inspection of the foregoing construction clearly shows that for $H =
\{e\}$ this definition effectively reduces to that of a
conventional particle field (cf. Corollary~\ref{cor:4.5}), i.e. a
section of the associated bundle $P \times_G {\mathbb V}$ defined by
a principal $G$-bundle and a representation $\rho: G \to
Aut({\mathbb V})$. Also, note that if $H \neq \{e\}$, then
${\mathcal V}_\rho({\mathcal P})$ is \textit{not} a vector space.
Nevertheless, one can define covariant derivatives on ${\mathcal
V}_\rho({\mathcal P})$, as is shown by the ensuing Examples 2 and 3.

Let $\{A_\alpha\}_{\alpha \in \Lambda}$ be the local potentials of a
connection ${\mathcal A}$ with respect to some presentation
${\mathcal T}$, as in Definition 6. Consider a point $x \in
U_\alpha$, and choose local coordinates $\{x^\mu\}$ in a
neighbourhood $U_x \subset U_\alpha$ of $x$ so that $A_\alpha =
A_{\alpha\mu}\rd x^\mu$ with $
A_{\alpha\mu} \in {\mathit\Gamma}(U_x, {\mathfrak g})$. Let ${\bar
v}_\alpha \in {\mathcal V}(U_\alpha)$ and choose a section $v_\alpha
= v_\alpha(y)$ representing ${\bar v}_\alpha$ in some neighbourhood
$W_x$ of $x$, with $W_x \subset U_x$. Then define
\begin{eqnarray}
[(\partial_\mu - A_{\alpha\mu}){\bar v}_\alpha](x) = ((\partial_\mu
- A_{\alpha\mu})v_\alpha)_x, \label{eq:7.2}
\end{eqnarray}
where $A_{\alpha\mu}$ acts upon $v_\alpha$ in accordance with the
Lie algebra representation induced by $\rho$. This germ is clearly
independent of the choice of $v$, and thus one obtains a
well-defined section of $(\partial_\mu - A_{\alpha\mu}){\bar
v}_\alpha \in {\boldsymbol\Gamma}(W_x,{\mathcal V})$. Moreover, for $x
\in U_\alpha \cap U_\beta$,
\begin{eqnarray}
(\partial_\mu - A_{\alpha\mu}(x)){\bar v}_\alpha(x) = {\bar
g}_{\alpha\beta}(x)(\partial_\mu - A_{\beta\mu}(x)){\bar v}_\beta(x),
\label{eq:7.3}
\end{eqnarray}
that is, if ${\bar v}_\alpha$ and ${\bar v}_\beta$ are related by
(\ref{eq:7.1}), then so are their covariant derivatives. To
establish (\ref{eq:7.3}) note that both sides of the equation are
well-defined local sections of ${\mathcal V}(U_\alpha \cap
U_\beta)$. Hence we may choose arbitrary representatives $v_\alpha$,
$v_\beta$, and $g_{\alpha\beta}$ of ${\bar v}_\alpha$, ${\bar
v}_\beta$, and ${\bar g}_{\alpha\beta}$, respectively, to perform
the calculation. We choose representatives satisfying the relation
$v_\alpha(x) = \rho(g_{\alpha\beta}(x))v_\beta(x)$, and by a routine
computation, as for conventional minimal coupling, we verify that
\begin{eqnarray}
(\partial_\mu - A_{\alpha\mu}(x))v_\alpha(x) = g_{\alpha\beta}(x)
(\partial_\mu - A_{\beta\mu}(x))v_\beta(x), \label{eq:7.4}
\end{eqnarray}
which implies (\ref{eq:7.3}). Choosing a sufficiently fine open
covering, we may, by paracompactness, identify the aforesaid open
sets $W_x$ with the $U_\alpha$, and hence assume that $(\partial_\mu
- A_{\alpha\mu}) {\bar v}_\alpha \in {\boldsymbol\Gamma}(U_\alpha,
{\mathcal V})$. Accordingly, we let $(\partial_\mu -
A_{\alpha\mu}){\bar v}_\alpha$ denote the element of ${\mathcal
V}({\mathcal P})$ determined by (\ref{eq:7.3}). For each $\alpha$
one can define a 1-form
\begin{eqnarray}
D^A_\alpha := (\partial_\mu - A_{\alpha\mu}) {\bar v}_\alpha \rd
x^\mu  \label{eq:7.5}
\end{eqnarray}
on $U_\alpha$ assuming values in the sheaf ${\mathcal V}$, and
(\ref{eq:7.3}) shows that these combine to provide a global
${\mathcal V}$-valued form $D^A {\bar v}$ on $X$.

The quantisation of the gauge and particle fields described above
will be investigated elsewhere. Accordingly, we shall not attempt to
define Lagrangians in the present paper. Rather, we shall merely
postulate that the equations of motion derived in conventional gauge
field theory are also valid in the present context. \vspace{0.2cm}

\textit{Example 2}. {\em Generalised interaction of Dirac monopole
with charged scalar field}. Consider the fundamental representation
of $U(1)$, that is, $\rho(\re^{{\rm i}\theta})z = \re^{{\rm
i}\theta} z$ for $z \in {\mathds C}$. We consider the Dirac sheaf
bundle ${\mathcal D}_\nu$ and connection\footnote{As explained in
\S\ref{sec:8} below, $A_\nu = \nu A$, where $A$ is the connection
form for the conventional Dirac monopole.} $A_\nu$ defined over the
subset $\tilde{\mathds R}^4$ of Minkowski space, as described in
\S6. Let ${\bar v} \in {\mathcal V}({\mathcal D}_\nu)$ be a particle
field associated with ${\mathcal D}_\nu$ by the representation
$\rho$, as in Definition~\ref{def:7}. The classical equation of
motion for a spin-0 charged particle interacting with an
electromagnetic potential $A$ on Minkowski space is
\begin{eqnarray}
(\partial_\mu - A_\mu)(\partial^\mu - A^\mu)\phi + m^2 \phi = 0,
\label{eq:7.6}
\end{eqnarray}
where the charge factor $\ri e$ has been absorbed into $A$. If
$\phi$ is replaced by ${\bar v} \in {\mathcal V}({\mathcal
D}_\nu)$, then, by virtue of relation (\ref{eq:7.3}), both sides
of the equation
\begin{eqnarray}
(\partial_\mu - A_\mu)(\partial^\mu - A^\mu) {\bar v} = - m^2
{\bar v} \label{eq:7.7}
\end{eqnarray}
are well-defined elements of ${\mathcal V}({\mathcal D}_\nu)$.
\vspace{0.2cm}

Remark: A solution of (\ref{eq:7.7}) might be physically described
as a `wave function' determined only up to a locally constant phase
factor. Current dogma holds that a globally constant phase factor is
undetectable, but the detectability of a locally constant phase
factor using a physical measuring apparatus also seems
problematical. Hence, the model described in the present example
appears to be physically plausible.

The use of induced bundles to describe conventional interactions
between gauge fields and particle fields initially defined in terms
of different principal bundles (see Bleecker 1981) can be extended
to $G_H$-bundles in a straightforward manner. If ${\mathcal F}$ and
${\mathcal F}'$ are, respectively, $G_H$- and $G'_{H'}$-bundles
defined over the same base space $X$, then, passing to a common
refinement if necessary, we may assume that ${\mathcal F}$ and
${\mathcal F}'$ are given in terms of presentations ${\mathcal T}$
and ${\mathcal T}'$ over the same system of trivialising
neighbourhoods $\{U_\alpha\}_{\alpha \in \Lambda}$ with transition
isomorphisms determined by ${\bar g}_{\alpha\beta}$ and ${\bar
g}'_{\alpha\beta}$, respectively. Then, the pairs $({\bar
g}_{\alpha\beta},{\bar g}'_{\alpha\beta})$ define a $(G\times
G')_{H\times H'}$-bundle ${\mathcal F}\times {\mathcal F}'$ over
$X$, with a presentation ${\mathcal T}\times {\mathcal T}' =
\{U_\alpha, {\mathcal F}_\alpha \times {\mathcal F}', {\bar
g}_{\alpha\beta} \times {\bar g}'_{\alpha\beta}\}$. Furthermore, if
$A_\alpha \in \Lambda^1(U_\alpha,{\mathfrak g})$ and $A'_\alpha \in
\Lambda^{\prime1}(U_\alpha,{\mathfrak g})$ are the local 1-forms of
connections ${\mathcal A}$ and ${\mathcal A}'$ relative to the
presentations ${\mathcal T}$ and ${\mathcal T}'$, respectively, then
the local 1-forms $A_\alpha\oplus A'_\alpha \in \Lambda^1(U_\alpha,
{\mathfrak g} \oplus {\mathfrak g}')$ determine a connection on
${\mathcal F} \times {\mathcal F}'$ relative to the presentation
${\mathcal T}\times{\mathcal T}'$. Now, let $\rho: G \to
Aut({\mathbb V})$ and $\rho': G' \to Aut({\mathbb V})$ be linear
representations on a vector space ${\mathbb V}$ such that
$\rho(g)\rho'(g') = \rho'(g')\rho(g)$ for all $(g,g') \in G \times
G'$. Then $(g,g') \mapsto \rho(g)\rho'(g')$ defines a linear
representation $\rho \times \rho': G \times G' \to Aut({\mathbb
V})$. The elements of ${\mathcal V}_{\rho\times\rho'}({\mathbb V})$
represent particle fields which interact with potentials defined on
${\mathcal F}$ as well as those defined on ${\mathcal F}'$.
\vspace{0.2cm}

\textit{Example 3}. {\em Generalised interaction of Dirac monopole
and spinor field}. Let $\rho$ denote the representation of $U(1)$ on
${\mathds C}^4$ defined by scalar multiplication $\rho(\re^{{\rm
i}\theta})\psi = \re^{{\rm i}\theta}\psi$, and let $\rho'=
D^{1/2,0}\oplus D^{0,1/2}:\, SL(2,{\mathds C}) \to Aut({\mathds C}^2
\times {\mathds C}^2) = Aut({\mathds C}^4)$, where $D^{1/2,0}(g)=g$
and $D^{0,1/2}(g)=(g^\dagger)^{-1}$. Then $\rho$ and $\rho'$
obviously commute in the foregoing sense, hence $(\rho \times
\rho')(\re^{{\rm i}\theta},g)\psi = \re^{{\rm i}\theta}\rho'(g)\psi$
defines a representation of $U(1) \times SL(2,{\mathds C})$ on
${\mathds C}^4$. Let ${\mathcal D}_\nu$ be the Dirac
$U(1)_{U(1)}$-bundle over ${\tilde{\mathds R}}^4$, with connection
forms $A_{\nu\alpha}$ ($\alpha= 1,2$) relative to the presentation
${\mathcal T}^\nu$, as in (\ref{eq:6.7}), extended from $S^2$ to
${\tilde{\mathds R}}^4$ in the obvious manner, as described in
\S\ref{sec:6}. Let ${\mathcal F}'$ be the trivial $SL(2,{\mathds
C})_{\{e\}}$-bundle over ${\tilde{\mathds R}}^4$, or essentially
${\mathcal F}' = {\mathcal G}^{SL(2,{\mathds C}
)}_{\{e\}}({\tilde{\mathds R}}^4)$, with trivial connection forms
$A'_\alpha=0$ corresponding to the trivial presentation ${\mathcal
T}' = \{{\tilde{\mathds R}}^4,{\mathcal F}'\}$. Thus, in this case
the local 1-forms $A_\alpha\oplus A'_\alpha = A_{\nu\alpha}\oplus 0$
so the action of $A_\alpha\oplus A'_\alpha$ upon the local section
$\bar{v}_\alpha \in {\mathcal V}(U_\alpha)$ is simply pointwise
multiplication by the imaginary number $A_{\nu\alpha}(x)$. The
conventional Dirac equation in the presence of an electromagnetic
potential, transcribed in the present context, becomes
\begin{eqnarray}
\gamma^\mu(\partial_\mu - A_{\nu\mu}) {\bar v} = -\ri m {\bar v}.
\label{eq:7.8}
\end{eqnarray}
Again, by virtue of (\ref{eq:7.3}), both sides of (\ref{eq:7.8})
represent well-defined elements of ${\mathcal
V}_{\rho\times\rho'}({\mathcal D}_\nu \times{\mathcal F}')$.

\section{Sheaf bundles for electromagnetic instantons}
\label{sec:8}

As indicated by the remarks in \S\ref{sec:1}, the mechanism of
charge quantisation \textit{a la} Dirac is not an exclusive feature
of the classical Dirac monopole alone. The essential ingredient is a
\textit{nontrivial} $U(1)$ bundle over some spacetime.\footnote{For
a trivial bundle, one need not satisfy any consistency (gauge
invariance) relations between local connection forms, hence, the
connection on an associated vector bundle could be any real multiple
$eA$.} Therefore, consideration of various other examples could be
theoretically instructive as well as suggesting possible examples of
applications to physical phenomena.

The foregoing Ansatz for construction of the Dirac sheaf bundle can
likewise be applied to cases where the underlying manifold $X$ is
not a two-sphere. Recall that given a principal bundle $P$ with
abelian structure group $G$ and transition functions
$g_{\alpha\beta}$ for some trivialising atlas ${\mathcal U}$, the
powers $g^n_{\alpha\beta}=(g_{\alpha\beta})^n$ for integral $n$ are
the transition functions of a principal $G$-bundle $P_n$ with
respect to the same atlas. Moreover, if $\{A_\alpha\}$ are the local
1-forms of a connection $A$ on $P$, then one can easily check that
$A_{n\alpha}:=nA_\alpha$ provide the local 1-forms of a connection
$A_n$ on $P_n$. Furthermore, if $F = \rd A$, then $F_n = \rd A_n =
nF$ for the curvature (field strength) of $A_n$. This means that if
$I(F^j)$ is a characteristic class of $P$, then the corresponding
class of $P_n$ is just $I(F_n^j) = n^j I(F^j)$. In particular, this
applies to the Chen classes.

Now, in accordance with the discussion in \S\ref{sec:6}, we can
proceed similarly in the context of $U(1)_{U(1)}$-bundles, replacing
the integer $n$ by the arbitrary real number $\nu$, and deduce that
if $c_j$ is the $j$th Chen class of the circle bundle $P$, then
$\nu^j c_j$ is the corresponding Chen class of the
$U(1)_{U(1)}$-bundle ${\mathcal P}_\nu$. To illustrate, consider the
canonical connection on a circle bundle over the complex projective
plane, which defines a gravitational and electromagnetic instanton.
As local minima of the Riemannian Hilbert-Maxwell action, such
instantons provide significant contributions to the partition
function in the joint path integral quantisation of the
gravitational and electromagnetic fields, thus playing a role
analogous to that of the usual instantons in the pure Yang-Mills
theory (cf. Gibbons \& Hawking 1979; Eguchi \& Hanson 1979).
\vspace{0.2cm}

\textit{Example 4}. \textit{Gravitational and electromagnetic
instanton}. Let $X=\mathds{CP}^2$ and consider the canonical $U(1)$
bundle over $X$. The complex projective plane $X= \mathds{CP}^2$
with coordinates $\{z_i\}_{i=1,2,3}$ satisfying ${\bar z}_j z^j=1$
is the quotient space of the five-sphere $S^5$ by the circle action
$z \to\re^{{\rm i}\phi}z$. We regard $S^5$ as a subspace of
${\mathds C}^3$:
\begin{eqnarray}
S^5 = \left\{ (z_1,z_2,z_3)\in{\mathds C}^3\big|\,
|z_1|^2+|z_2|^2+|z_3|^2= 1 \right\} . \label{eq:8.1}
\end{eqnarray}
The Hopf map $\pi: S^5\to\mathds{CP}^2$ is defined by
$\pi(z_1,z_2,z_3) = (\bar{z}_3z_1+\bar{z}_1z_3,\, \ri\bar{z}_3
z_1-\ri\bar{z}_1z_3,\, \bar{z}_2z_3+\bar{z}_3z_2, -\ri\bar{z}_2z_3 +
\ri \bar{z}_3z_2,\, \bar{z}_3z_3 -\bar{z}_2z_2-\bar{z}_1z_1)$.
Parameterising $S^5$ by
\begin{eqnarray}
z_1=\sin\half\theta_1 \cos\half\theta_2 \re^{{\rm i}\phi_1}, \quad\!
\! z_2=\sin\half\theta_1 \sin\half\theta_2 \re^{{\rm i}\phi_2},
\quad \!\! z_3=\cos\half\theta_1 \re^{{\rm i}\phi_3}, \label{eq:8.3}
\end{eqnarray}
where $\theta_1,\theta_2\in[0,\pi]$, $\phi_i\in{\mathds R}$, formula
(8.2) becomes
\begin{eqnarray}
\pi(z_1,z_2,z_3) &=& \left(\sin\theta_1\cos\half\theta_2
\cos(\phi_3-\phi_1) , \sin\theta_1\cos\half\theta_2
\sin(\phi_3-\phi_1), \right. \nonumber
\\ && \hspace{-2.0cm} \left. \sin\theta_1\sin\half\theta_2
\cos(\phi_3-\phi_2), \sin\theta_1\sin\half\theta_2
\sin(\phi_3-\phi_2), \cos\theta_1 \right). \label{eq:8.4}
\end{eqnarray}
As trivialising neighbourhoods, let $U_\alpha = \mathds{CP}^2 -
\{z_\alpha=0\}$. Writing $(\zeta^1,\zeta^2)=(z_1/z_3,z_2/z_3)$ for
the inhomogeneous coordinates on $U_3$, a section $\sigma_3$ over
$U_3$ is given by
\begin{eqnarray}
\sigma_3 = \frac{1}{\sqrt{1+|\zeta^1|^2+|\zeta^2|^2}} \left(
\begin{array}{c} \zeta^1 \\  \zeta^2 \\ 1 \end{array} \right)
= \left( \begin{array}{l} \sin\half\theta_1 \cos\half\theta_2
\re^{{\rm i}\varphi_1} \\  \sin\half\theta_1 \sin\half\theta_2
\re^{-{\rm i}\varphi_2} \\ \cos\half\theta_1
\end{array} \right). \label{eq:8.5}
\end{eqnarray}
Similarly, we have
\begin{eqnarray}
\sigma_1 = \left( \begin{array}{l} \sin\half\theta_1
\cos\half\theta_2
\\  \sin\half\theta_1 \sin\half\theta_2 \re^{{\rm i}\varphi_3} \\
\cos\half\theta_1 \re^{-{\rm i}\varphi_1} \end{array} \right), \quad
\sigma_2 = \left( \begin{array}{l} \sin\half\theta_1
\cos\half\theta_2 \re^{-{\rm i}\varphi_3} \\ \sin\half\theta_1
\sin\half\theta_2
\\ \cos\half\theta_1 \re^{{\rm i}\varphi_2} \end{array} \right),
\label{eq:8.6}
\end{eqnarray}
for sections over $U_1$ and $U_2$, respectively. By virtue of the
relation $\varphi_1+\varphi_2+\varphi_3=0$ the transition function
of the $U(1)$ bundle on, say, $U_1\cap U_3 \subset \mathds{CP}^2$ is
given by $\sigma_3=\re^{ {\rm i}\varphi_1} \sigma_1$. Similarly, we
have $\sigma_2=\re^{{\rm i}\varphi_2}\sigma_3$ and $\sigma_1=\re^{
{\rm i}\varphi_3}\sigma_2$. Thus the generic transition function is
simply $g_{\alpha\beta}=\re^{{\rm i}\varphi}$. The canonical
connection on the bundle is given by the Hermitian inner product
$\omega=\langle\bar{z},\rd z\rangle$. In terms of the local
coordinates on $U_3$, given by
$\zeta^1=\tan\half\theta_1\cos\half\theta_2\re^{{\rm i}\varphi_1}$
and $\zeta^2=\tan\half\theta_1\sin\half\theta_2 \re^{-{\rm
i}\varphi_2}$ the local connection form is $\omega_3 = \ri [
\sin^2\half\theta_1\cos^2\half\theta_2 \rd \varphi_1 -
\sin^2\half\theta_1\sin^2\half\theta_2 \rd\varphi_2]$. A similar
calculation shows that the local connection form on
$\{U_\alpha\}_{\alpha=1,2}$ reads $\omega_1 = \ri [
\sin^2\half\theta_1\sin^2\half\theta_2 \rd \varphi_3 -
\cos^2\half\theta_1 \rd\varphi_1 ]$ and $\omega_2 = \ri [
\cos^2\half\theta_1 \rd \varphi_2 - \sin^2 \half\theta_1
\cos^2\half\theta_2 \rd\varphi_3]$. Using the relation
$\varphi_1+\varphi_2 +\varphi_3=0$, we see at once that the local
connection forms are related by the gauge transformation
$\omega_1=\omega_2+g_{12}^{-1}\rd g_{12}$, where $g_{12}=\re^{{\rm
i}\varphi_3}$. Similarly we have $\omega_2=\omega_3+g_{23}^{-1}\rd
g_{23}$ with $g_{23}=\re^{{\rm i}\varphi_2}$ and $\omega_3=\omega_1+
g_{31}^{-1}\rd g_{31}$ with $g_{31}=\re^{{\rm i} \varphi_1}$.
Calculating the field strength $F=\rd\omega$ on $U_3$, say, we
obtain
\begin{eqnarray}
F &=& \ri \left[ \sin\theta_1\, \rd\theta_1 \wedge
\left(\cos^2\half\theta_2\, \rd \varphi_1 - \sin^2\half\theta_2\,
\rd \varphi_2\right) \right. \nonumber \\ && \left. ~~ - \sin^2
\half\theta_1 \sin\theta_2\, \rd\theta_2 \wedge (\rd\varphi_1
+\rd\varphi_2) \right]. \label{eq:8.9}
\end{eqnarray}
This expression for $F$ in (\ref{eq:8.9}) agrees with the one
obtained by Trautman (1977); an alternative expression is given by
Gibbons \& Pope (1978) using different local coordinates. The field
$F$ is self-dual with vanishing energy-momentum tensor, and thus,
along with the Fubini-Study metric on $\mathds{CP}^2$, solves the
Einstein-Maxwell equation with cosmological constant. At this point,
one can merely speculate upon the possible incorporation of the
corresponding $U(1)_{U(1)}$-bundles ${\mathcal P}_\nu$ into the
above-mentioned path integral formalism. \vspace{0.2cm}

\textit{Example 5}. The field $F$ defined on $\mathds{CP}^n$ induces
solutions to the Maxwell equation on analytic submanifolds of
$\mathds{CP}^n$. Trautman (1977) considered an example given by the
Veronese embedding (cf. Brody \& Hughston 2001) of $\mathds{CP}^1$
in $\mathds{CP}^n$. For $n = 2$ this is the embedding
$(z_1,z_2)\hookrightarrow(z_1^2, \sqrt{2}z_1z_2,z_2^2)$, which
defines a conic ${\mathcal C}$ in $\mathds{CP}^2$. A short
calculation shows that, in terms of the spherical polar coordinates
$(\theta,\phi)$ of $\mathds{CP}^1\simeq S^2$, the local connection
forms of the bundle on the two hemispheres
$\{U_\alpha\}_{\alpha=1,2}$ are given by $\omega_\alpha=\ri
((-1)^{\alpha+1}-\cos\theta) \rd\phi$. Comparison with
(\ref{eq:6.7}) for $\nu=1$ shows that $\omega_\alpha =2A_\alpha$.
Thus, the electromagnetic field induced on $S^2$ corresponds to a
magnetic pole of unit strength, which might appropriately be called
the \textit{Trautman monopole}. Another elementary example is the
solution to the Maxwell equation arising from the Segr\'e embedding
of $\mathds{CP}^1\times \mathds{CP}^1$ in $\mathds{CP}^3$; this
defines a quadric ${\mathcal Q}$ in $\mathds{CP}^3$. In terms of the
spherical polar coordinates $(\theta_1,\theta_2,\phi_1,\phi_2)$ of
${\mathcal Q}$ one finds at once that the local connection forms on
${\mathcal Q}$ are given by $\omega_\alpha=\half\ri\left[
((-1)^{\alpha+1} -\cos\theta_1) \rd\phi_1+ ((-1)^{\alpha+1}
-\cos\theta_2)\rd\phi_2 \right]$ in the respective trivialising
neighbourhoods $\{U_\alpha\}_{\alpha=1,2}$. Thus, we obtain a pair
of disjoint Dirac monopoles. The potential physical significance of
the extensions to nonintegral charges $\nu$ would, of course, be
similar to that of the simple Dirac sheaf bundles described in
\S\ref{sec:5}. \vspace{0.2cm}

Remark: A nontrivial example of a new solution to the Maxwell
equation on a torus $T^2$ can be constructed by pulling the field
$F$ back to $T^2$ via the elliptic curve $(z)\hookrightarrow
(1,\wp(z),\wp'(z))$ in $\mathds{CP}^2$. Here, $z$ is the complex
coordinate of the torus and $\wp(z)$ denotes the Weierstra{\ss}
$\wp$-function.

Remark: Since the equivalence classes of principal $U(1)$-bundles
over $X$ are indexed by $H^2(X,{\mathds Z})$ (see Lawson \&
Michelsohn 1990, Appendix A), any realisation of the usual
hypothesis of charge quantisation through a topological mechanism is
excluded in cosmological models over a contractible base, such as
the Schwarzschild universe (see also Trautman 1979). This
constitutes an apparent contradiction with certain grand unified
theories which require the existence of monopoles ('t Hooft 1974).
The author is not aware of any Ansatz whereby this apparent
contradiction could be resolved.

\section{Quasispin structures and Dirac operators}
\label{sec:9}

Consider the short exact sequence of Lie groups
\[
\begin{CD}
1  @>>> H @>>> G @>\phi>> K @>>> 1
\end{CD}
\]
with $H$ closed and central in $G$. As described in Example 1, a
principal $G$-bundle $(P,\pi,X)$ canonically determines a
$G_H$-bundle ${\mathcal F}_H(P)$ over $X$. In the special case where
$H$ is discrete, a principal $K$-bundle $(Q, \pi,X)$ canonically
determines a $G_H$-bundle over $X$ as follows. Let
$\{k_{\alpha\beta}\}$ denote the transition functions of $Q$
relative to a trivialising open covering ${\mathcal U} =
\{U_\alpha\}_{\alpha \in \Lambda}$ of $X$. Since $H$ is discrete,
the above epimorphism $\phi:G \to K$ is a local homeomorphism,
hence, for any $x \in U_\alpha \cap U_\beta$, there exists a
neighbourhood $V_x$ of $x$, with $V_x \subset U_\alpha \cap
U_\beta$, and a smooth mapping $g_{\alpha\beta}^x:V_x \to G$ such
that $\phi \circ g_{\alpha\beta}^{x} = k_{\alpha\beta}|_{V_x}$, and
any two such $g_{\alpha\beta}^{x}$ and ${g}_{\alpha\beta}^{\prime x}$
are related by $g_{\alpha\beta}^{x}=h {g}_{\alpha\beta}^{\prime
x}$ for some constant $h \in H$. Therefore, $g_{\alpha\beta}^{x}$
uniquely defines an element ${\tilde g}_{\alpha\beta}^{x}$ of
${\mathfrak G}^G_H(V_x)$ (see (\ref{eq:3.5})) and hence a
continuous section ${\bar g}^x_{\alpha\beta}$ of the sheaf
${\mathcal G}^G_H(V_x)$. For $y \in V_x \cap U_\alpha \cap U_\beta
\cap U_\gamma$ we have
\begin{eqnarray}
\phi(g_{\alpha\beta}^{x} (y) g_{\beta\gamma}^{x}(y)
g_{\gamma\alpha}^{x} (y)) = k_{\alpha\beta}(y) k_{\beta\gamma}(y)
k_{\gamma\alpha}(y) = 1, \label{eq:9.1}
\end{eqnarray}
hence, $g_{\alpha\beta}^{x}(y) g_{\beta\gamma}^{x}(y)
g_{\gamma\alpha}^{x} (y) \in {\rm Ker}(\phi) = H$, and since $H$ is
discrete, this product must be constant in some neighbourhood $W_x
\subset V_x$ of $x$. Thus,
\begin{eqnarray}
{\tilde g}_{\alpha\beta}^{x} {\tilde g}_{\beta\gamma}^{x} {\tilde
g}_{\gamma\alpha}^{x} = 1 \label{eq:9.2}
\end{eqnarray}
in $G_H(W_x)$, and we replace $V_x$ by $W_x$. Passing to a
suitable refinement of ${\mathcal U}$, we may, by paracompactness,
identify the $W_x$ with the intersections $U_\alpha \cap U_\beta$,
discard the superscripts $x$ from the ${\tilde
g}_{\alpha\beta}^{x}$, and thereby obtain, as in Example 1,
transition sheaf isomorphisms
\begin{eqnarray}
T_{\alpha\beta}: {\mathcal F}_\alpha|_{U_\alpha \cap U_\beta} \to
{\mathcal F}_\beta|_{U_\alpha \cap U_\beta} , \label{eq:9.3}
\end{eqnarray}
where ${\mathcal F}_\alpha := {\mathcal G}_H(U_\alpha)$, which satisfy
the cocycle condition by virtue of (\ref{eq:9.2}). We denote the
resulting $G_H$-bundle by ${\mathcal S}_Q(H)$.

In particular, if $P$ is a principal $G$-bundle covering $Q$ in
the sense that the transition functions $g_{\alpha\beta}$ of $P$
with respect to some common trivialising open covering for $P$ and
$Q$ satisfy the relation
\begin{eqnarray}
\phi(g_{\alpha\beta})(x) = k_{\alpha\beta}(x), \label{eq:9.4}
\end{eqnarray}
then we may select $g_{\alpha\beta}^{x} = g_{\alpha\beta}(x)$ in the
foregoing construction, and thus ${\mathcal S}_Q(H)$ coincides with
the $G_H$-bundle ${\mathcal F}_P(H)$ described in Example 1.

Let ${\mathfrak k}$ denote the Lie algebra of $K$, and let $A$ be a
conventional connection on $Q$. Relative to a given system of local
trivialisations over the open covering ${\mathcal U}= \{U_\alpha
\}_{\alpha \in \Lambda}$, with transition functions
$k_{\alpha\beta}$, $A$ is determined by a family of ${\mathfrak
k}$-valued 1-forms $A_\alpha \in \Lambda^1(U_\alpha,{\mathfrak k})$
satisfying
\begin{eqnarray}
A_\alpha(x) = k_{\alpha\beta}(x)A_\beta(x)k_{\beta\alpha}(x) +
k_{\alpha\beta}(x)\rd k_{\beta\alpha}(x) \label{eq:9.5}
\end{eqnarray}
for all $x \in U_\alpha \cap U_\beta$. Relative to another system of
local trivialisations over ${\mathcal U}$, with transition functions
$k'_{\alpha\beta}(x) = k_\alpha^{-1}(x) k_{\alpha\beta}(x)
k_\beta(x)$, the same connection $A$ is represented by the 1-forms
\begin{eqnarray}
A'_\alpha(x) = k_\alpha^{-1}(x) A_\alpha(x) k_\alpha(x) +
k_\alpha^{-1}(x) \rd k_\alpha(x).   \label{eq:9.6}
\end{eqnarray}
For a sufficiently fine covering, we can choose sections $g_\alpha
\in {\mathit\Gamma}(U_\alpha,G)$ and $g_{\alpha\beta} \in
{\mathit\Gamma}(U_\alpha \cap U_\beta, G)$ such that
$\phi(g_\alpha(x)) = k_\alpha(x)$ and, as above,
$\phi(g_{\alpha\beta}(x)) = k_{\alpha\beta}(x)$. These choices
are unique modulo constant multiplicative factors in
${\mathit\Gamma}_c(U_\alpha,H)$ and ${\mathit\Gamma}_c(U_\alpha \cap
U_\beta,H)$, respectively. Furthermore, the covering map $\phi$
induces a Lie algebra isomophism $\phi_*:\, {\mathfrak g} \to
{\mathfrak k}$. Applying $\phi_*^{-1}$ to (\ref{eq:9.5}) and
(\ref{eq:9.6}), we obtain
\begin{eqnarray}
\phi_*^{-1} A_\alpha(x) = g_{\alpha\beta}(x)(\phi_*^{-1} A_\beta(x))
g_{\beta\alpha}(x) + g_{\alpha\beta}(x) \rd g_{\beta\alpha}(x)
\label{eq:9.7}
\end{eqnarray}
and
\begin{eqnarray}
\phi_*^{-1}A'_\alpha(x) = g_\alpha^{-1}(x) (\phi_*^{-1}
A_\alpha(x))g_\alpha(x) + g_\alpha^{-1}(x) \rd g_\alpha(x).
\label{eq:9.8}
\end{eqnarray}
Let ${\bar g}_{\alpha\beta}$ and ${\bar g}_\alpha$ denote the
sections of ${\mathcal G}^G_H(U_\alpha\cap U_\beta)$ and ${\mathcal
G}^G_H(U_\alpha)$ determined by $g_{\alpha\beta}$ and $g_\alpha$,
respectively, and write $\phi_*^{-1} A_\alpha = {\tilde A}_\alpha$,
$\phi_*^{-1} A'_\alpha = {\tilde A}'_\alpha$. Then (\ref{eq:9.7})
and (\ref{eq:9.8}) imply the relations
\begin{eqnarray}
{\tilde A}_\alpha(x) = {\bar g}_{\alpha\beta}(x) {\tilde A}_\beta(x)
{\bar g}_{\beta\alpha}(x) + {\bar g}_{\alpha\beta}(x) \rd {\bar
g}_{\beta\alpha}(x)  \label{eq:9.9}
\end{eqnarray}
and
\begin{eqnarray}
{\tilde A}'_\alpha(x) = {\bar g}_{\alpha}^{-1}(x) {\tilde
A}_\alpha(x) {\bar g}_{\alpha}(x) + {\bar g}_{\alpha}(x) \rd {\bar
g}_{\alpha}(x),  \label{eq:9.10}
\end{eqnarray}
in accordance with the explanatory remarks following (\ref{eq:6.1})
and (\ref{eq:6.3}). Thus, the family of 1-forms $\{{\tilde
A}_\alpha\}_{\alpha \in \Lambda}$, corresponding to the presentation
${\mathcal T}=\{U_\alpha, {\mathcal F}_\alpha, {\bar
g}_{\alpha\beta}\}_{\alpha, \beta\in \Lambda}$, where, as usual,
${\mathcal F}_\alpha={\mathcal G}^G_H(U_\alpha)$, determines a
connection ${\mathcal A}$ on the $G_H$-bundle ${\mathcal S}_Q(H)$ in
the sense of Definition~\ref{def:6}.

Of particular physical interest is the case where $G = Spin^e(r,
s)$, $K = SO^e(r, s)$ (the superscript $e$ denoting the component of
the identity) and $\phi$ is the canonical homomorphism with kernel
$H = {\mathds Z}_2$. Given a principal $K$-bundle $Q$ over $X$, a
well-known theorem (Lawson \& Michelsohn 1990, Friedrich 2000)
states that $Q$ is covered by some principal $G$-bundle $P$ in the
above sense iff the second Stiefel class $w_2(Q) \in H^2(X, {\mathds
Z}_2)$ vanishes, and then the equivalence classes (i.e. bundle
equivalence compatible with the covering map onto $Q$) of such
coverings $P$ (known as \textit{spin structures}) are parameterised
by the elements of $H^1(X, {\mathds Z}_2)$. Inequivalent spin
structures may (or may not) be inequivalent as abstract $G$-bundles
over $X$. Nevertheless, this can entail no inconsistency with
Proposition~\ref{prop:1}, since, if the hypothesis of that
proposition is satisfied, then, by the above-cited theorem, there
exists at most one spin structure $P$ over $Q$, and hence at most
one ${\mathcal F}_P(H)$, which, if existent, coincides with
${\mathcal S}_Q(H)$. However, the \textit{quasispin structure}
${\mathcal S}_Q(H)$ exists in any case, even if ${\mathcal F}_P(H)$
does not.

Specifically, let $X$ be a smooth Lorentzian manifold and $Q$ the
orthonormal frame bundle of $X$. The Levi-Civita connection $A_{LC}$
on $Q$ then determines, as above, a connection ${\mathcal A}_{LC}$
on ${\mathcal S}_Q(H)$, and the spin representation $\rho: G \to
Aut({\mathbb S})$, where ${\mathbb S}$ denotes the spin module,
gives rise to a space ${\mathcal V}:= {\mathcal V}_\rho({\mathcal
S}_Q(H))$ of particle fields (see Definition~\ref{def:7}). Covariant
differentiation in ${\mathcal V}$ is then well-defined, just as was
indicated, using local coordinates, in Examples 2 and 3.

Relative to a given trivialisation over $U_\alpha$ in a sufficiently
fine open covering, a quasispinor field $\psi \in {\mathcal V}$ is
represented by a smooth section $\psi_\alpha \in
{\mathit\Gamma}(U_\alpha,{\mathbb S})$, determined up to a constant
factor $h
\in H$, that is, up to a sign $\pm 1$. Using the local 1-form
$(A_{LC})_\alpha$ relative to the given trivialisation, one obtains
a conventional Dirac operator $D\!\!\!\!/_\alpha$ over each
$U_\alpha$, which, applied to $\psi_\alpha$, yields a section
$D\!\!\!\!/_\alpha \psi_\alpha$, likewise defined up to a sign. Let
${\bar\psi}_\alpha$ denote the element of ${\mathcal V}(U_\alpha)$
determined by $\psi_\alpha$. Since the Dirac operator behaves
covariantly with respect to gauge transformations $g \in G$, the
relations (\ref{eq:7.1}), with ${\bar v}_\alpha =
{\bar\psi}_\alpha$, imply corresponding relations for the sections
$D\!\!\!\!/_\alpha \psi_\alpha$, which thus combine to yield an
element ${\mathcal D}\!\!\!\!/_\alpha \psi \in {\mathcal
V}_\rho({\mathcal S}_Q(H))$.

According to a well-known theorem of Geroch (1968), the orthonormal
frame bundle of a noncompact Lorentzian manifold $X$ possesses a
spin structure iff $X$ is parallelisable. A specific example of a
noncompact Lorentzian manifold not admitting a spin structure is
described, e.g., in Clarke (1971). The present formalism permits the
global description of fermionic fields on such spacetimes, and the
local properties of quasispinor fields and quasi-Dirac operators
in this context are clearly identical with those of their
conventional counterparts. The foregoing considerations are likewise
applicable to the construction of quasispinor fields and
quasi-Dirac operators on oriented Riemannian manifolds, with $G =
Spin(n)$, $K = SO(n)$ and $H = {\mathds Z}_2$. In the conventional
theory, the index of the Dirac operator on a `spinnable' manifold
$X$ is expressible, by the Atiyah-Singer index formula, in terms of
the Chen classes of $X$ (Lawson \& Michelsohn 1990; Berline
\textit{et al}. 1992). Even if $X$ is not spinnable, this expression
is still well-defined. On the other hand, since ${\mathcal V}$ is
not a vector space, the index of the quasi-Dirac operator
${\mathcal D}\!\!\!\!/$ cannot be defined in the conventional
manner. Of course, one could, by fiat, simply define the index of
the quasi-Dirac operator as the value of the Atiyah-Singer
expression. However, a more interesting approach would consist in
providing a general geometrical definition of the index for
differential operators on ${\mathcal V}$, and proving that its value
for the quasi-Dirac operator is given by the Atiyah-Singer formula.
This appears to pose a not entirely trivial problem.

In the pseudoriemannian case, if $X$ is not space and time
orientable, then one deals with pin structures (see Chamblin and
Gibbons 1997 for a discussion of physical applications). Since the
homomorphism $Pin(r,s) \to O(r,s)$ is also a double covering, the
foregoing construction is directly applicable, and one thus obtains
\textit{quasipin structures}, even if true pin structures fail to
exist.

$Spin^c$-structures can also be generalised in a similar vein, but
since in this case the central subgroup $H = U(1)$ is not discrete,
the situation is somewhat more complicated and will be dealt with
elsewhere.

\section{Discussion}
\label{sec:10}

Mathematically sophisticated expositions of gauge theory tend to
define connections and curvature \textit{ab initio} on the total
space of a principal bundle before pulling these quantities down to
the base space via local trivialisations. Some readers may, perhaps,
inquire why we have not adopted this more elegant approach. The
reason is that our sheaf bundles are not Hausdorff spaces, hence not
manifolds, so the tangent spaces are undefined. Consequently, we
must adopt the orthodox physical practice of working in terms of
local sections.

In the relevant literature one occasionally observes heuristic
statements to the effect that topological solitons already display
certain quantal effects at the classical level. Since charge
renormalisation and running coupling constants can only result from
field quantisation in the conventional theory, one could, in a
similar vein, assert that the generalised solitons described above
already display quantum field theoretic effects at the classical
level. The coupling constants can walk even before they begin to
run.

We have constructed a family of generalised Dirac monopoles
${\mathcal D}_\nu$ which include the conventional ${\mathcal D}_n$
as special cases. The ${\mathcal D}_n$ interact with conventional
wave functions $\psi$, i.e. sections of the associated vector
bundles, and likewise, mathematical consistency in the description
of interactions involving the ${\mathcal D}_\nu$ requires the
introduction of generalised wave functions ${\mathit\Psi}$ which are
sections of certain sheaves. Every conventional $\psi$ defines a
${\mathit\Psi}$, but there also exist ${\mathit\Psi}$ that do not
arise from any $\psi$. For example, consider normalised wave
functions defined over the circle $X = S^1$. In the conventional
model, these include, e.g., $\psi(x) = \exp(\ri nx)$ for integral
$n$, but not $\exp(\ri \nu x)$ for arbitrary real $\nu$. In our
model, however, the $\exp(\ri\nu x)$ (and more generally,
$\exp(\ri\nu x)f(x)$, where $f(x)$ is an ordinary normalised
complex-valued function on $S^1$) represent well-defined sections
of the sheaf ${\mathcal G}^G_H(S^1)$, where $G=H=U(1)$. Thus, rather
than demanding knowledge of a global function on $X$, determined
only up to a globally constant phase factor, we merely demand, for
any point $x \in X$, the knowledge of a function on some
neighbourhood $U$ of $x$, and determined only up to a constant phase
factor on $U$. The probabilistic interpretation of such a
generalised wave function remains the same as in the conventional
theory. Moreover, differentiation with respect to $x$ and
multiplication by ordinary functions $V(x)$ are well-defined, so we
can apply Hamiltonian operators such as $H=-\rd^2/\rd x^2+V(x)$.

Magnetic monopoles, of either integral charge or otherwise, have not
yet been encountered in experimental studies (see, e.g., Milton 2006
for the current status of the experimental limits). The discovery of
either nonintegral magnetic monopoles or nonintegral electric
charges would tend to indicate that the above-described model is not
merely more general mathematically but also more realistic
physically than the conventional principal bundle model. Such a
discovery might therefore cast doubt upon the long-standing
assumption that the venues for physical phenomena are necessarily
manifolds, and suggest that sheaves, which provide a more local
description, constitute the appropriate arenas for physical models.

The sheaf-theoretic dequantisation of solitons and instantons for
nonabelian gauge groups constitutes a challenging topic for
research, and might have implications for grand unified theories.
The above treatment of the Dirac sheaf bundle can be adapted to
cases where the Dirac monopole is embedded within a nonabelian
principal bundle, such as the 't Hooft monopole.

The validity of the model constructed above would have certain
obvious physical consequences. Firstly, the detection of a magnetic
monopole would not necessarily imply the quantisation of
electromagnetic charges. Conversely, an observed violation of charge
quantisation would not necessarily imply the nonexistence of
magnetic monopoles.

A definite conclusion concerning the actual physical quantisation of
electric charges remains elusive. For example, explosions of
electron bubbles in liquid helium have been observed in the
laboratory, suggesting that fractionally charged particles can
possibly exist in isolation (Konstantinov \& Maris 2003), although
the interpretation of such experiments is controversial (Jackiw
\textit{et al}. 2001; Bender \textit{et al}. 2005). Other physical
observations apparently consistent with nonintegral charges include
the fractional quantum Hall effect (Laughlin 1983) and geometric
phase measurements in anisotropic spin systems (Bruno 2004). The
possible variability of the fine structure constant over the
cosmological timescale (Bekenstein 1982) and the postulated
fractional charges of the quarks in the deconfinement phase could
also be relevant to this issue. A sheaf-theoretic formulation of the
underlying quantum theory along the foregoing lines might
conceivably clarify some of these diverse phenomena.

\begin{acknowledgements}
The author hereby expresses his gratitude to C.~M.~Bender,
E.~J.~Brody, G.~W.~Gibbons, T.~W.~B.~Kibble, B.~Muratori, and
R.~P.~W.~Thomas for stimulating and informative discussions in
connection with the foregoing material.
\end{acknowledgements}

\end{document}